\documentclass[preprint,1p]{elsarticle}

\usepackage{amssymb}
\usepackage{eurosym} % per il simbolo dell'euro
\usepackage{mathrsfs}  %per gli integrali funzionali

\DeclareRobustCommand{\Eqref}[1]{Eq.~(\ref{#1})}
\DeclareRobustCommand{\Tabref}[1]{Tab.~(\ref{#1})}
\DeclareRobustCommand{\Figref}[1]{Fig.~(\ref{#1})}
\DeclareRobustCommand{\Secref}[1]{Sec.~(\ref{#1})}

\journal{}

\def\eg{\emph{e.g.} }
\def\ie{\emph{i.e.} }

\begin{document}

\begin{frontmatter}


\title{QCD simulations with staggered fermions on GPUs}

\author[pisa]{C.~Bonati}
\ead{bonati@df.unipi.it}

\author[kek]{G.~Cossu}
\ead{cossu@post.kek.jp}

\author[genova]{M.~D'Elia}
\ead{Massimo.Delia@ge.infn.it}

\author[pisa]{P.~Incardona}
\ead{asmprog32@hotmail.com}

\address[pisa]{Dipartimento di Fisica, Universit\`a di Pisa and INFN, Largo Pontecorvo 3, I-56127 Pisa, Italy}
\address[kek]{KEK IPNS, Theory Center, 1-1 Oho, Tsukuba-shi, Ibaraki 305-0801, Japan}
\address[genova]{Dipartimento di Fisica, Universit\`a di Genova and INFN, Via Dodecaneso 33, I-16146 Genova, Italy}

\begin{abstract}

We report on our implementation of the RHMC algorithm for the simulation of lattice QCD with two 
staggered flavors on Graphics Processing Units, using the 
NVIDIA CUDA programming language. The main feature of our code is that the GPU is not 
used just as an accelerator, but instead the whole Molecular Dynamics trajectory 
is performed on it. After pointing out the main bottlenecks and how to 
circumvent them, we discuss the obtained performances. We present some preliminary results
regarding OpenCL and multiGPU extensions of our code and discuss future perspectives.
\end{abstract}


\end{frontmatter}

\section{Introduction}

Graphics processing units (GPUs) have been developed originally as 
co-processors meant to fast deal with graphics tasks.
In recent years the video game market developments compelled GPUs 
manufacturers to increase the floating point calculation performance 
of their products, by far exceeding the performance of standard CPUs 
in floating point calculations. 
The architecture evolved toward programmable many-core chips that are 
designed to process in parallel massive amounts of data.  
These developments suggested the possibility of using GPUs in the field 
of high-performance computing (HPC) 
as low-cost substitutes of more traditional 
CPU-based architectures: nowadays such possibility is being fully exploited
and GPUs represent an ongoing breakthrough for many computationally
demanding scientific fields, providing consistent computing resources 
at relatively low cost, also in terms of power consumption (watts/flops).

Due to their many-core architectures, with fast access to the on-board memory,
GPUs are ideally suited for numerical tasks allowing for data parallelism,
\ie for SIMD (Single Instruction Multiple Data) parallelization.
The numerical simulation, by Monte Carlo algorithms, of the
path integral formulation of Quantum Field Theories, discretized on a
Euclidean space-time lattice, is a typical such task: one has to sample
the distribution for a system with many degrees of freedom and mostly
local interactions. 

Quantum Chromodynamics (QCD), the Quantum Gauge Theory describing strong 
interactions, is a typical example where numerical simulations represent
the best tool to investigate systematically specific features of the theory
and give an answer to many important unsolved questions, regarding
e.g. color confinement, deconfinement and the values of hadron masses.
Lattice QCD and its computational needs has represented a challenge 
in the field of HPC since more than 30 years, being often the stimulus
for new machine developments (think e.g. of the series of APE machines~\cite{ape}).

The introduction of GPUs in lattice QCD calculations started with the seminal work of 
Ref.~\cite{videogame}, in which the native graphics APIs were used, but the real explosion of 
interest in the field followed the introduction of NVIDIA's CUDA (Compute Unified Device 
Architecture) platform, that effectively disclosed the field of GPGPU (General Purpose GPU 
\cite{cuda}), providing a more friendly programming environment. 

GPUs have maintained their role of co-processors in most numerical
applications, where they are used as accelerators for some 
specific, time demanding purposes.
In the same spirit, most of previous studies 
on the application of GPUs to lattice QCD 
calculations were mainly aimed at using them together 
with the standard architectures in order to speed up 
some specific steps, typically the expensive Dirac matrix 
inversion. Our intent is instead to use GPUs in substitution of the 
usual architectures, actually performing the whole 
simulation by them: one still needs a CPU to run the main 
program, but mostly in the role of a mere controller of the GPU
instruction flow.

To achieve this result we found simpler to 
write a complete program from scratch instead of using 
existing software packages\footnote{On earlier stage we wrote 
a staggered version of JLab's  Chroma working on GPUs.}, 
in order to have a better control of all the steps to be 
performed and ultimately transferred 
to the GPU. Our implementation uses NVIDIA's CUDA platform 
together with a standard C++ serial control program 
running on CPU. The specific case considered in the present
study regards QCD on a hypercubic lattice with
quark fields discretized in the standard staggered
formulation.

The paper is organized as follows. In Section~\ref{simalg} we
give more details about the lattice discretization of QCD 
considered in our study and the simulation algorithm adopted.
In Section~\ref{gpufeatures} we review some of the fundamental 
features of GPU architectures. In Section~\ref{implem} we describe 
our implementation of the algorithm on GPUs and discuss the achieved
performances. Finally, in Sections~\ref{opencl} and~\ref{parallel}
we discuss some preliminary comparisons with performances obtained
with OpenCL and multiGPU implementations of our code. A preliminary
report about our implementation has been presented in Ref.~\cite{gpuproc}.

\section{Lattice QCD and the simulation algorithm}
\label{simalg}

QCD is a Quantum Field Theory based on the symmetry under
local non-Abelian gauge (color) transformations belonging to  
the group of special unitary $3 \times 3$ complex
matrices, $SU(3)$. It describes six different flavors
of spin 1/2 colored particles (quarks), which transform in 
the fundamental (triplet) representation of $SU(3)$ and interact
through the gauge field, which lives in the Algebra of the 
color gauge group and describes 8 colored, spin 1 particles 
known as gluons.

An elegant, gauge invariant lattice 
discretization of the theory is given in terms
of gauge link variables $U_\mu (n)$, where
$n$ indicates a lattice site and $\mu$ is one
of the four Euclidean space-time directions \cite{Wilson74}.
They are the elementary
parallel transporters belonging to the gauge group and 
associated to each elementary link connecting two neighboring
sites of the lattice. Hence in total we have 
$4 L_x L_y L_z L_t$ $SU(3)$ matrices, 
where $L_\mu$ is the number of lattice sites along direction
$\mu$, which in present simulations is typically not larger than
$10^2$.
Quark fields $\psi (n)$ instead live on lattice sites
and carry a color index, hence they are complex color triplets, 
one for each flavor and Dirac index.

The discretized Euclidean Feynman path integral, giving e.g. a 
representation of the thermal partition function, is written
as 
\begin{eqnarray}\label{pathint}
Z=\int \mathscr{D}U \mathscr{D}\bar{\psi}\mathscr{D}\psi e^{-S_g[U]-\bar{\psi}M[U]\psi}
\end{eqnarray}
where $S_g$ is the pure gauge part of the action, describing 
gluon-gluon interactions and written in terms
of traces of products of link variables over closed loops,
while 
$\bar{\psi}M[U]\psi$ is a bilinear form in the fermionic variables,
which describes quark-gluon interactions, with $M[U]$ a large sparse
matrix written in terms of the gauge link variables.

The functional integration in Eq.~(\ref{pathint}) is over all link
variables (gauge group invariant integration for each link) and
all quark fields. Actually,
due to their fermionic nature, the quark fields appearing
in the path integral are Grassmann anticommuting variables; the best
way we know to numerically deal with them is to integrate
them explicitly. That results in the appearance of the determinant
of the fermion matrix $M[U]$:
\begin{eqnarray}\label{determinant}
Z=\int \mathscr{D}U \mathscr{D}\bar{\psi}\mathscr{D}\psi e^{-S_g[U]-\bar{\psi}M[U]\psi}
\propto \int \mathscr{D}U \det(M[U]) e^{-S_g[U]} \; .
\end{eqnarray}
Notice that, in general, a fermion determinant appears for each quark
species and that the determinant term
becomes a trivial constant when all quarks have
infinite mass and decouple (pure gauge or quenched limit). 
One can show that,
apart from specific difficult cases (e.g. QCD at finite baryon density), 
the integrand in 
Eq.~(\ref{determinant}) is a positive quantity, admitting a probabilistic
interpretation, so that one can 
approach the numerical computation of the path integral by 
importance sampling methods, 
which are typically based on dynamic Monte Carlo algorithms.

The most difficult, time consuming part in such algorithms 
consists in taking properly into account the fermion determinant
$\det(M[U])$.
The best available 
method is to introduce dummy bosonic complex fields $\phi$, 
which come in the same number as the original fermionic variables
and are known as pseudo-fermions~\cite{WP}:
\begin{eqnarray}\label{pseudofermions}
Z \propto \int \mathscr{D}U \left( \det(M[U]) \right)^2 e^{-S_g[U]}
\propto \int \mathscr{D}U \mathscr{D}\phi \, \exp\Big(-S_g[U]-\phi^* \big(M[U]^{\dag}M[U]\big)^{-1}\phi\Big)
\end{eqnarray}
where we have explicitly considered the case of two identical quark species,
as in the case of two light flavors ($u$ and $d$ quarks) with all 
other flavors decoupled.

The particular lattice discretization implemented in the present study
considers a simple plaquette action for the pure gauge term, i.e. 
products of four gauge link variables around the elementary closed
square loops (plaquettes) of the lattice, and a standard staggered 
discretization for the fermionic term. That means that fermion fields 
living on lattice sites have only color indexes (Dirac indexes can be 
reconstructed afterwards combining fields living on different lattice sites),
while the fermionic matrix reads as follows:
\begin{eqnarray}
\label{stag}
M_{n_1,n_2} [B,q] = a m \delta_{n_1,n_2} 
+ {1 \over 2} \sum_{\nu=1}^{4}\eta_\nu (n_1) \left(
 U_\nu (n_1) 
\delta_{n_1,n_2-\hat\nu} - 
U^{\dag}_\nu (n_1-\hat\nu) \delta_{n_1,n_2+\hat\nu} 
\right) \:,
\end{eqnarray}
where $n_1$ and $n_2$ are 4-vectors with integer components labeling
lattice sites, $\hat\nu$ is an elementary lattice versor, 
$\eta_\nu(n) \equiv (-1)^{n_x + n_y + \dots n_{\nu-1}}$ are
known as staggered fermion phases and $a$ is the lattice spacing.
Color indexes are implicit in Eq.~(\ref{stag}) (the identity in color 
space is understood for the mass term proportional to $a m$).

The staggered discretization differs from other (e.g. Wilson) fermion
discretizations for the absence of the additional Dirac index: that
implies lighter algebra computations which have an effect both on the 
overall computational complexity and on the maximal performances,
as we shall explain in details later on.

A particular feature of the staggered fermion matrix in Eq.~(\ref{stag}) is
that it describes four flavors. When simulating a different number
of flavors one has to use a trick known as rooting. $N_f$ flavors 
of equal mass are described by the following partition function
\begin{eqnarray}\label{rooting}
Z \propto \int \mathscr{D}U \left( \det(M[U]) \right)^{N_f/4} e^{-S_g[U]}
\, .
\end{eqnarray}

\subsection{Numerical algorithm}

A convenient algorithm to simulate the action in \Eqref{pseudofermions} is the Hybrid Monte 
Carlo~\cite{hmc} (HMC). The idea is very simple and it is conveniently exposed by using as an 
example the case of a single variable with action $S=V(\varphi)$, i.e. distributed
proportionally to $\exp (-V(\varphi)) d\varphi$. 
As a first step a dummy variable $p$, corresponding to a momentum
conjugate to $\varphi$, is introduced using the following 
identity:
\begin{eqnarray}
\label{hmcsimple}
\int d\varphi \exp (-V(\varphi)) \propto 
\int d\varphi  d p \exp \left(-V(\varphi) - \frac{1}{2} p^2 \right) \, .
\end{eqnarray}
It is trivial that expectation values over $\varphi$ are untouched by the
introduction of $p$, which is a stochastically
independent variable. The basic idea of the HMC algorithm 
is to sample the distribution in $p$ and $\varphi$ by first 
extracting a value of $p$ according to its Gaussian distribution
and then performing a joint molecular dynamics evolution of $p$ and $\varphi$
which keeps the total ``energy'' $V(\varphi) + p^2/2$ unchanged, 
obtaining an updated value of $\varphi$ as a final result. Going into
more details, the HMC proceeds as follows:
\begin{enumerate}
\item a random initial momentum is generated with probability $\propto e^{-\frac{1}{2}p^2}$;
\item starting from the state $(\varphi,p)$, a new trial state $(\varphi',p')$ is generated by 
numerically solving in the fictitious time $\tau$ the equations of motion derived from the 
action $V(\phi)+\frac{1}{2}p^2$, \ie
\begin{eqnarray}
\dot q = p \; ; \qquad
\dot p = - \frac{\mathrm{d} V}{\mathrm{d} \varphi} \; ;
\end{eqnarray}
such equations are integrated numerically using a finite time
step $\Delta \tau$. As a consequence the energy is conserved only up
to some power of $\Delta \tau$, depending on the 
integration scheme adopted;
\item the new state $(\varphi',p')$ is accepted with probability $\mathrm{min}(1,e^{-\delta S})$ where
$\delta S=S(\varphi',p')-S(\varphi,p)$ (Metropolis step).
\end{enumerate}
It can be shown (see \eg \cite{hmc, Kennedy}) that the sequence of the $\varphi$ configurations obtained in this way is 
distributed with the correct $e^{-V(\varphi)}$ probability provided the solution of the equation of motion satisfies the
requirements
\begin{itemize}
\item the evolution is reversible, \ie 
\begin{equation}
(\varphi,p)\to (\varphi',p') \Leftrightarrow (\varphi',-p')\to (\varphi,-p)
\end{equation}
\item the evolution preserves the measure of the phase space, \ie 
\begin{equation}
\det\frac{\partial (\varphi',p')}{\partial (\varphi,p)}=1
\end{equation}
\end{itemize}
A large class of integrators that satisfy these two constraints are the so-called symmetric symplectic
integrators, the simplest member of this class being the leap-frog or $PQP$ scheme (for improved schemes see \eg
\cite{CP, CG, TF}).

In the particular case of the action in \Eqref{pseudofermions}, the momenta are conjugate
to the gauge link variables: they are therefore $3 \times 3$ complex matrices $H_\mu (n)$ 
(one for each gauge link) living in the algebra of the gauge group, i.e. they are 
traceless hermitian matrices writable as $H_\mu (n) = \sum_a T^a \omega_\mu^a(n)$ where
$T^a$ are the group generators, and the action associated with momenta is simply
given by $\sum_{n,\mu} {\rm Tr} (H_\mu (n) H_\mu (n))$.
A convenient implementation of the picture above is then given by 
the so-called $\Phi$ algorithm of Ref.~\cite{Phi}:
\begin{enumerate}
\item a vector $R$ of complex Gaussian random numbers is generated and 
the pseudofermion field is initialized by $\phi=M[U]^{\dag }R$, in such a way that the probability distribution for
$\phi$ is proportional to $\exp(-\phi^* (M^{\dag}M)^{-1}\phi)$;
\item the momenta are initialized by Gaussian random matrices (i.e. each 
$\omega^a_\mu(n)$ is extracted as a normally distributed variable);
\item the gauge field and momenta are updated by using the equations of motion;
\item the final value of the action is computed and the Metropolis step performed.
\end{enumerate}
Point 3 is the more time consuming, since the calculation of the force requires at each step to solve the 
sparse linear system 
\begin{equation}\label{system}
\Big(M[U]^{\dag}M[U]\Big)X=\phi
\end{equation}
which is usually performed by means of Krylov methods (see \eg \cite{Vorst}). For staggered fermions a complication 
is the presence of the $4-$th root of the determinant in the action: \Eqref{system} becomes
\begin{equation}\label{system_r}
\Big(M[U]^{\dag}M[U]\Big)^{N_f/4}X=\phi
\end{equation}
where $N_f$ is the number of degenerate flavors. In order to overcome this problem the Rational Hybrid Monte Carlo (RHMC) 
was introduced in \cite{rhmc}, in which the root of the fermion matrix is approximated by a rational function, which is then 
efficiently computed by means of the shifted versions of the Krylov solvers (see \eg \cite{shifted}). 

In order to speed-up the simulations, the following common tricks were implemented
\begin{itemize}
\item even/odd preconditioning \cite{DGR}
\item multi-step integrator, with action divided in gauge and fermion part \cite{SW}
\item improved integrator, second order minimum norm, see \eg \cite{TF} 
\item multiple pseudo-fermions to reduce the fermion force magnitude and increase integration step size \cite{CK}
\item different rational approximations and stopping residuals for the Metropolis and the Molecular Dynamic steps \cite{CK}
\end{itemize}

\section{Fundamental NVIDIA GPU architecture features}
\label{gpufeatures}

\begin{figure}
\centering
\scalebox{0.2}{\rotatebox{0}{\includegraphics*{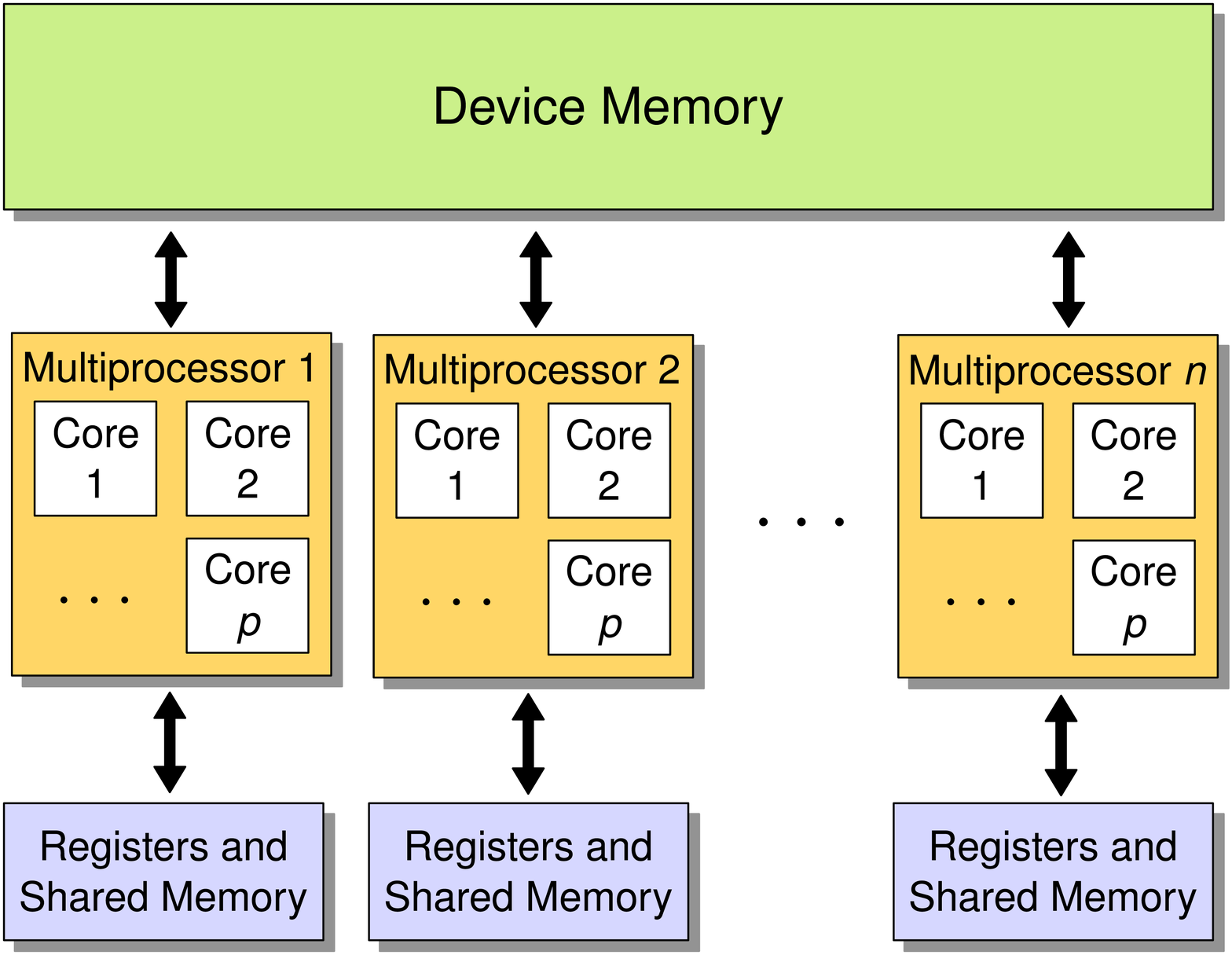}}}
\caption{Architecture of a modern NVIDIA graphics card. Figure taken from \cite{cuda_fig}.}
\label{arch_fig}
\end{figure}

In this section we will review the main features of the GPUs architecture which are to be taken into account in
order to efficiently use their computational capabilities. 
Modern GPUs are massively parallel computing elements, composed of hundreds of cores grouped into multiprocessors. 
The typical architecture of a modern NVIDIA graphic card is outlined in \Figref{arch_fig} and the most important point 
for the following is the presence of three different storage levels. 
Roughly speaking the architecture of ATI cards is similar.

Primary storage is provided by the device memory, which is accessible by all multiprocessors but has a relatively 
high latency. Within the same multiprocessor, cores have also access to local registers and to shared memory.
Shared memory is accessible by the threads of the same multiprocessor and its access is orders of 
magnitude faster than device memory one, being very close to the computing units; however, while the total amount of device 
memory is of order of few GBs, the local storage is only 16KB per multiprocessor both for the registers and for the shared 
memory\footnote{For NVIDIA Tesla cards 10 series. The 20 series has 64 KB of on-chip memory that can be partitioned 
as shared and L1 cache.}, so that it is typically impossible to use just these local fast memories. 

In order to hide the latency time of the device memory it is convenient to have a large number of threads in concurrent 
execution, so that when data are needed from device memory for some threads, the ones ready to execute are immediately 
sent to computation. The highest bandwidth from device memory is achieved when a group of 16 threads accesses a contiguous 
memory region (coalesced memory access), because its execution requires just one instruction call, saving a lot of 
clock-cycles. This will be crucial in the following, when discussing the storage model for the gauge configuration.

Double precision capability was introduced with NVIDIA's GT200 generation, the first one specifically designed having 
in mind HPC market, and by now there is only a factor $2$ between the peak performance in single and double precision. 
In \Tabref{spec_tab} the specifications of the GPUs used in this work are reported. 

\begin{table}[t]
\centering
\begin{tabular}{|l|l|l|l|l|l|}
\hline 
GPU  & Cores & Bandwidth & Gflops  & Gflops  & Device   \\
     &       & (GB/s)    & single  & double  & memory   \\
     &       &           & (peak)  & (peak)  & (GB)     \\
\hline\hline
Tesla C1060 & 240 & 102 & 933 & 78 & 4 \\
\hline 
Tesla C2050/2070 & 448 & 144 & 1030 & 515 & 3/6 \\
\hline
\end{tabular}
\caption{Specifications of the NVIDIA cards used in this work.}\label{spec_tab}
\end{table}

Communications between the GPU and the CPU host are settled by a PCI-E bus, whose typical bandwidth is 5GB/s, 
to be compared with the GPU internal bandwidth between device memory and cores of order 100 GB/s. This is clearly 
the main bottleneck in most of GPU applications. When allowed by memory constraints the optimal strategy is thus 
to copy the starting gauge configuration (and momenta) on the device at the beginning of the simulation and to 
perform the complete update on the GPU, instead of using it just to speed up some functions and transferring 
gauge field back and forth between host and device memories. With two flavors of fermions we checked that the 
largest lattice fitting on the device memory of a C1060 card is about a $32^4$ one, which is too small for typical 
zero temperature calculations but large enough for finite temperature ones, for which the temporal extent of the  
lattice is typically much smaller than the spatial one (lattices as large as 
a $50^3 \times 8$ or $64^3 \times 4$ fit well on the same card).

In our implementation of the Dirac kernel a different thread is associated with every even\footnote{Because of the odd/even 
preconditioning pseudo-fermions are defined on even sites only.} site in the fermion update 
and to every link in the gauge update, so that different threads do not cooperate. Shared memory is thus used just as 
a local fast memory and, unfortunately, no data reuse is possible. This setup is forced by the high ratio between 
data transfer and floating point operations, which is
around 1.5 bytes/flop for the single precision Dirac kernel.

\section{Details of the implementation}
\label{implem}

In the following we shall discuss various aspects of our implementation
of the RHMC algorithm on GPUs and present the achieved performances.
Our guiding strategy has been that of bringing as much as possible
of the computations on the GPU, leaving for the CPU only light or 
control tasks: such strategy has the twofold advantage 
of exploiting the computational power of the GPU at its best
and of minimizing costly memory transfers between the CPU
and the GPU; the strategy is facilitated by the fact that all heavy tasks of a 
lattice QCD computation can be easily parallelized.

\begin{figure}
\centering
\scalebox{0.45}{\rotatebox{0}{\includegraphics*{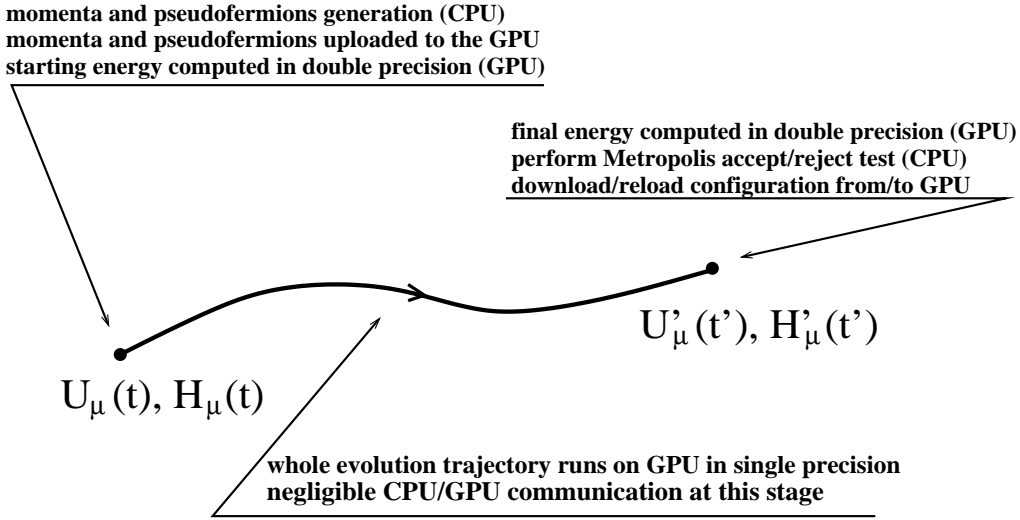}}}
\caption{Sketch of our implementation of the RHMC algorithm on GPUs}
\vspace{1cm}
\label{traj_fig}
\end{figure}

A sketch of our typical implementation of the RHMC trajectory is reported
in Fig.~\ref{traj_fig}. The heat-bath creation of momenta and 
pseudofermions at the beginning of the trajectory is the only 
part of the code which, even if portable to the GPU, has been kept on
the CPU: we have decided to do so since it 
is computationally very light and since in this way we have 
avoided to have a random number generator running on the GPU.
The whole molecular dynamics part is run on the GPU in single precision,
with negligible involvement of the CPU.

\subsection{Precision issues}

We will now address the issues related to the use of double precision. The main drawback of double precision is clear 
from Tab.~\ref{spec_tab}: single precision floating point arithmetic always outperforms the double one, although in 
the Fermi architecture the double precision penalty was significantly reduced. Another motivation to prefer the 
single precision is to speed up internal memory transfers because lattice QCD calculations are typically bandwidth limited.

The need for double precision is related to the evaluation of the action for the Metropolis step, to be 
performed at the end of a molecular dynamics trajectory and which guarantees the stochastic correctness of the 
RHMC algorithm (see also \Secref{inv_sec}). Because of that the first and the last Dirac inversions (the ones needed for 
the Metropolis) are performed in double precision, while the inversions needed in the fermion force calculation are in 
single precision. The update of the gauge field is always performed in single precision and double precision is used only 
in the reunitarization. 

In order that the HMC algorithm reproduces the correct probability distribution, i.e. that it respects the detailed balance principle,
 another property which has to be ensured is the reversibility of the 
molecular dynamics trajectories~\cite{hmc}. Since the gauge updates use only single precision we can expect reversibility 
to be valid only up to single precision; this will be extensively analyzed  in the following, see \Secref{rev_test_sec}.

\begin{figure}
\centering
\setlength{\unitlength}{1cm}
\begin{picture}(10,2.1)

\put(0,0.7){\line(1,0){10}} 
\put(0,1.4){\line(1,0){10}}
\put(0,2.1){\line(1,0){10}} 
\put(0,2.8){\line(1,0){10}}  

\put(0,0.7){\line(0,1){2.1}}   
\put(1,0.7){\line(0,1){2.1}}   
\put(2,0.7){\line(0,1){2.1}}   
\put(3,0.7){\line(0,1){2.1}}   
\put(4,0.7){\line(0,1){2.1}}   
\put(5,0.7){\line(0,1){2.1}}   
\put(6,0.7){\line(0,1){2.1}}   
\put(7,0.7){\line(0,1){2.1}}   
\put(8,0.7){\line(0,1){2.1}}   
\put(9,0.7){\line(0,1){2.1}}   
\put(10,0.7){\line(0,1){2.1}}  

\footnotesize
\put(0.1,2.35){$S_{1}(1)$}
\put(1.1,2.35){$S_{1}(2)$} 
\put(2.1,2.35){$S_{1}(3)$} 
\put(3.3,2.35){$\cdots$} 
\put(4.3,2.35){$\cdots$} 
\put(5.1,2.35){$S_{2}(1)$}
\put(6.1,2.35){$S_{2}(2)$} 
\put(7.1,2.35){$S_{2}(3)$} 
\put(8.3,2.35){$\cdots$} 
\put(9.3,2.35){$\cdots$} 

\put(0.3,1.65){$\cdots$} 
\put(1.1,1.65){$S_{3}(1)$}
\put(2.1,1.65){$S_{3}(2)$} 
\put(3.1,1.65){$S_{3}(3)$} 
\put(4.3,1.65){$\cdots$} 
\put(5.3,1.65){$\cdots$} 
\put(6.1,1.65){$D_{1}(1)$}
\put(7.1,1.65){$D_{1}(2)$} 
\put(8.1,1.65){$D_{1}(3)$}
\put(9.3,1.65){$\cdots$} 

\put(0.1,0.95){$D_{2}(1)$}
\put(1.1,0.95){$D_{2}(2)$} 
\put(2.1,0.95){$D_{2}(3)$} 
\put(3.3,0.95){$\cdots$} 
\put(4.3,0.95){$\cdots$} 
\put(5.1,0.95){$D_{3}(1)$}
\put(6.1,0.95){$D_{3}(2)$} 
\put(7.1,0.95){$D_{3}(3)$} 
\put(8.3,0.95){$\cdots$} 
\put(9.3,0.95){$\cdots$} 

\end{picture}
\caption{Gauge field storage model: $S_1(k), S_2(k), S_3(k)$ are three \texttt{float4} that store the $32$ most 
significant bits of the $k-$link's elements. Analogously $D_1(k), D_2(k), D_3(k)$ are \texttt{float4} that store 
the $32$ less significant bits.} 
\label{gauge_fig}
\end{figure}

\subsection{Memory allocation scheme}

As previously noted a correct allocation scheme is of the utmost importance in order to efficiently use the device 
memory. For the case of the staggered fermion discretization of QCD, the storage of the gauge configuration is the 
most expensive one, so we will concentrate just on this. Similar techniques can be used also for the storage of 
the momenta and the pseudo-fermions.

As stated before, QCD calculations on GPU are typically bandwidth limited. This can be easily seen by noting that 
in simulations the largest amount of time is needed by the Krylov linear solver, whose elementary step is the 
product between the Dirac matrix and the pseudo-fermion fields, which is essentially a huge number of products 
between $3\times 3$ unitary matrices and complex $3-$vectors. To compute every single product an equivalent of 72 
real number elementary operations have to be performed and 96 bytes of memory have to be allocated (in 
single precision). By using the specifications given in \Tabref{spec_tab} we then  see, \eg for a C1060 card, 
that the maximum performance achievable in single precision is below $10\%$ of the peak performance of the GPU. 

It is thus convenient not to storage all the elements of the $SU(3)$ matrices, but to use a representation in 
terms of fewer parameters: in this way we can reduce the amount of memory exchange at the expense of increasing 
the computational complexity. The additional calculations do not introduce significant overhead and they are 
actually negligible compared to the gain in the memory transfers. We used a 12 real number representation: only the 
first two rows of the $3\times 3$ unitary matrices are stored, while the third one is reconstructed on fly when needed.
It is actually possible to further reduce memory transfers by adopting a minimal 8 parameter representation
of $SU(3)$ matrices~\cite{clark}; however that requires considerable computational overhead which limits the 
additional gain obtained, therefore we decided to not implement it in our code.

Since in the Metropolis step the inversion of the Dirac matrix in double precision is required, we need to store a double 
precision gauge configuration, although in most of the calculations it will be used just as a single precision one.
In order not to waste bandwidth and device memory, it is useful to write a double precision number $a$ by using two
single precision numbers $b$ and $c$: $b$ is defined by the 32 most significant bits of $a$, while $c$ stores the
32 less significant ones. In C language this amounts to
\begin{eqnarray*}
&& b=\mathtt{(float)} a\\
&& c=\mathtt{(float)} (a-\mathtt{(double)}b)
\end{eqnarray*}
In computations where only single precision is required we can just use $b$ instead of $a$, otherwise there are 
two different strategies available: to use $b$ and $c$ directly, effectively avoiding the explicit use of double 
precision arithmetic (see \eg \cite{gst}) or to reconstruct the double precision number $a$ to be used in calculations. 
Although the first method allows for the use of hardware without double precision capabilities, we implemented this 
second method, which is expected to be more efficient on double precision capable hardware. 

To get coalesced memory accesses it is crucial for blocks of thread in execution to use contiguous regions of 
device memory. This behavior is maximized if we adopt the storage model shown in Fig.~\ref{gauge_fig}: the index in 
parenthesis identifies the link and range in the interval $[1, 4\times\mathrm{volume}]$, the most significant bits of 
the first two rows of the $k-$th $SU(3)$ link matrix are grouped in three \texttt{float4}, denoted by $S_1(k), S_2(k)$ 
and $S_3(k)$, analogously $D_1(k), D_2(k)$ and $D_3(k)$ take into account the less significant bits. The use of texture 
memory is a further improvement used to reduce the effects of the residual imperfect memory accesses.  

The performance of the Dirac operator kernel (one application of the matrix \Eqref{stag} to a random vector) in single 
precision which is obtained by means of this storage scheme is shown in Tab.~\ref{perf_tab}: while using $60-70\%$ of 
the bandwidth, only $5-6\%$ of the peak performance is reached, consistently with the previous analysis.

\begin{table}[t]
\centering
\begin{tabular}{|l|l|l|}
\hline
Lattice & Bandwidth GB/s & Gflops \\
\hline
$4\times 16^3$ & $56.84\pm 0.03$ & $49.31 \pm 0.02$ \\
\hline
$32\times 32^3$ & $64.091\pm 0.002$ & $55.597\pm 0.002$ \\
\hline
$4\times 48^3$ & $69.94\pm 0.02$ & $60.67\pm 0.02$ \\
\hline
\end{tabular}
\caption{Staggered Dirac operator kernel performance figures on a C1060 card (single precision).}\label{perf_tab}
\end{table}

\subsection{Inverter}\label{inv_sec}

As noted in \Secref{gpufeatures} it is convenient to execute on the GPU complete sections of the code instead of using it just to 
speed up some specific functions. The most time consuming of these sections is the inversion of the Dirac operator.

The inversion of the Dirac operator in lattice QCD simulations is usually performed by using Krylov space solvers; for
staggered fermions the optimal choice is the simplest one of this class of solvers: the Conjugate Gradient (CG) algorithm
(see \cite{dF}).

In all Krylov space solvers the approximate solution and its estimated residual are calculated recursively and the value
of the estimated residual is used to terminate the algorithm. While in exact arithmetic the estimated residual coincides 
with the residual of the approximate solution, in finite precision this is no more the case and the estimate diverges form
the true value. 

It can be shown that, for the solution of the linear equation $Ax=b$ by means of Krylov methods, the following 
estimate holds (see \cite{greenbaum})
\begin{equation}\label{th_prec}
\frac{\|b-Ax_{(k)}-r_{(k)}\|}{\|A\| \|x\|}\le \epsilon\, O(k)\left(1+\max_{j\le k}\frac{\|x_{(j)}\|}{\|x\|}\right)
\end{equation}
where $x_{(k)}$ is the approximate solution at the $k-$step of the algorithm, $r_{(k)}$ is its estimated residual and
$\epsilon$ is the machine precision.

For a single precision Dirac inversion a typical value for the minimum true residual is $10^{-2}-10^{-3}$: that is too
large to ensure the correctness of the Metropolis step, therefore at least in 
that case a double precision solver is thus needed.

In standard Krylov solvers this problem can be overcome by using the residual replacement strategy: sometimes the 
true residual is explicitly calculated in double precision and the algorithm is restarted. By this method it is 
possible to obtain reliable results, as happens with double precision calculations, but using almost always single precision 
arithmetics. Residual replacement methods are well understood theoretically \cite{vy} and have been successfully 
applied to QCD calculations on GPUs \cite{clark}. 

However, in RHMC we need solvers for shifted systems, i.e. for the system
\begin{equation}\label{shifted}
(A+\sigma_i)x=b
\end{equation}
for various $\sigma_i$ values. Krylov solvers for shifted systems exist and they allow to reuse the results of the
matrix products needed to solve $(A+\sigma_0)x=b$ to compute the solution also of $(A+\sigma_i)x=b$ for $i>0$
(see \eg \cite{shifted}). The algorithm of these solvers is however much more rigid than the usual one of Krylov solvers 
and in particular the starting guess solution has to be the null one, thus preventing the possibility of restarting. 
For this reason the Dirac inversions in the Metropolis step have to be performed completely in double 
precision. 

Recently it was noted in \cite{APGL} that, on GPUs, it could be most convenient to use ordinary Krylov solvers also
to compute the solutions of \Eqref{shifted}, in order to allow the use of the 
residual replacement strategy and of other techniques to improve the convergence of the solver, like preconditioning or 
multigrid, which are of difficult implementation for shifted solvers.

\subsection{Algorithm tests}\label{rev_test_sec}

We will now report on some test performed in order to gain a better understanding of the possible influence of the 
mixed precision setting on the simulation results.

The tests have been performed in the $N_f=2$ theory, with two degenerate quark flavors of bare mass $am=0.01335$, and mostly on finite temperature lattices 
with a time extension $N_t = 4$. With such settings
the deconfinement transition is known to be located at 
$\beta_c\approx 5.272$~\cite{nf2}, therefore we have chosen to work at
$\beta=5.264$ in order to be in the confined phase with broken chiral symmetry, where almost zero modes are expected to 
exist for the Dirac matrix, as a consequence of the Banks-Casher relation \cite{BC}.

Two lattices of extension $4\times 16^3$ and $4\times 32^3$ have been tested; in both cases the number of pseudofermions was $2$ 
and the rational approximations adopted in the Molecular Dynamic and Metropolis step were of degree $8$ and $15$ 
respectively. A statistics of
$O(10^3)$ MD trajectories, each of length $\tau = 1$, 
has been collected for each test. We denote by $n_{md}$ the number of (fermionic) 
integration step used in the simulation and $d\tau_f/d\tau_g$ the ratio between the fermionic integration step and the
gauge one used in the multi-step $2MN$ integrator. The parameters $r_{md}$ and $r_{metro}$ are the stopping residuals to
be used for the Krylov solver in the MD evolution and in the Metropolis step respectively.

To test the reversibility of the MD evolution, at the end of the trajectory ($\tau=1$) the sign of the momenta was reversed and the
evolution continued until $\tau=2$. We then measured the quantities (see \cite{inst})
\begin{eqnarray}
\Delta C=\sqrt{\sum_{t,x,y,z,\mu}\|U^{(\tau=0)}_{\mu}(t,x,y,z)-U^{(\tau=2)}_{\mu}(t,x,y,z)\|^2 / \mathrm{dof} }\\
\Delta M=\sqrt{\sum_{t,x,y,z,\mu}\|H^{(\tau=0)}_{\mu}(t,x,y,z)+H^{(\tau=2)}_{\mu}(t,x,y,z)\|^2 / \mathrm{dof} }
\end{eqnarray}
where \(\|\cdot\|\) is the matrix $2-$norm  ($\|A\|^2=\sum_{ij}|A_{ij}|^2$) and 
$\mathrm{dof}=4\times 8\times N_s^3\times N_t$  is the number of degrees of freedom. $\Delta C$ and $\Delta M$ are 
thus estimators of the reversibility violation for degree of freedom of the gauge fields and the momenta respectively.

We tested various combination of single and double precision inverters:
\begin{itemize}
\item[$D_1$:] Metropolis in double precision, $r_{metro}=10^{-9}$, MD in double precision, $r_{md}=10^{-9}$
\item[$D_2$:] Metropolis in double precision, $r_{metro}=10^{-7}$, MD in double precision, $r_{md}=10^{-7}$
\item[$D_3$:] Metropolis in double precision, $r_{metro}=10^{-5}$, MD in double precision, $r_{md}=10^{-5}$
\item[$D_4$:] Metropolis in double precision, $r_{metro}=10^{-3}$, MD in double precision, $r_{md}=10^{-3}$
\item[$F_1$:] Metropolis in single precision, $r_{metro}=10^{-3}$, MD in single precision, $r_{md}=10^{-3}$
\item[$MP_1$:] Metropolis in double precision, $r_{metro}=10^{-9}$, MD in single precision, $r_{md}=10^{-3}$
\item[$MP_2$:] Metropolis in double precision, $r_{metro}=10^{-7}$, MD in single precision, $r_{md}=10^{-3}$
\end{itemize}

\subsubsection{$4\times16^3$ lattice}

\begin{figure}[p]
\centering
\scalebox{0.4}{\rotatebox{0}{\includegraphics*{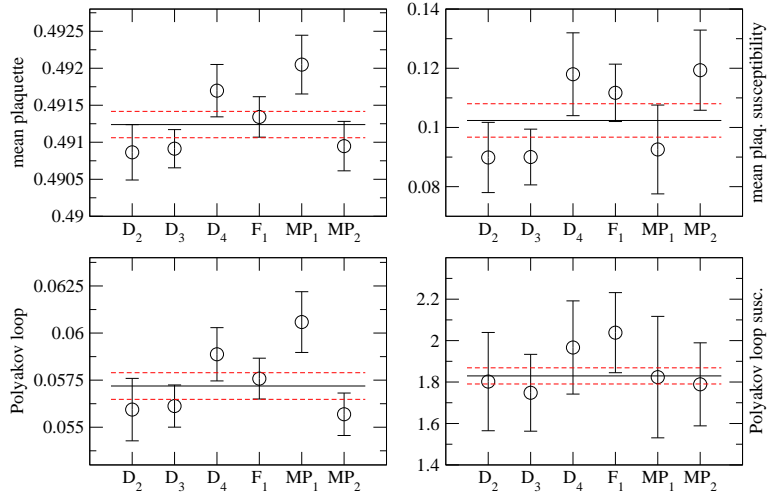}}}
\caption{Lattice $4\times 16^3$, values of some observables for the different runs. The black solid line is the result of
a fit on all the values, the red dashed are drowned one standard deviation away from the average.}\label{obs_value}
\end{figure}

For this lattice we used $n_{md}=12$ and $d\tau_f/d\tau_g=10$. All the different runs started from a common 
thermalized configuration and in \Figref{obs_value} the values of some 
observables are shown, from which it follows that all the different runs give compatible results.

\begin{figure}[p]
\centering
\scalebox{0.26}{\rotatebox{0}{\includegraphics*{4x16_5.264_rev_conf.eps}}}\hspace{0.5cm}
\scalebox{0.26}{\rotatebox{0}{\includegraphics*{4x16_5.264_rev_mom.eps}}}
\caption{Lattice $4\times 16^3$, values of the reversibility estimators.}\label{16_rev}
\end{figure}
\begin{figure}[p]
\centering
\scalebox{0.26}{\rotatebox{0}{\includegraphics*{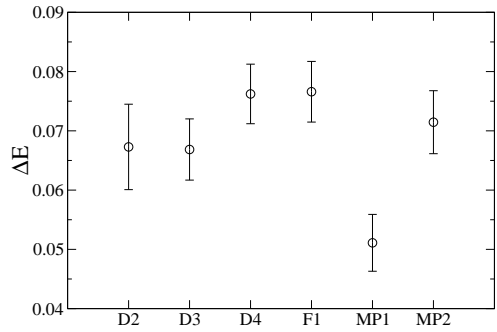}}}
\caption{Lattice $4\times 16^3$, difference between the final and the initial value of the action along an MD trajectory.}
\label{16_dE}
\end{figure}

The values of the reversibility estimators $\Delta C$ and $\Delta M$ are shown in \Figref{16_rev} and, as expected, 
they are compatible with reversibility violations at the level of single precision, which are inevitable since the 
gauge update is completely performed in single precision. In this setting the precision of the inversions does not seem to 
influence the reversibility of the algorithm in a sensible way: the differences between the various runs are of order
of $1\%$. Also the difference between the action at the beginning and end of an MD trajectory 
was monitored and its behavior is shown in \Figref{16_dE}; again no appreciable difference is 
observed between the different runs.

\begin{figure}[p]
\centering
\scalebox{0.26}{\rotatebox{0}{\includegraphics*{4x32_5.264_rev_conf.eps}}}\hspace{0.5cm}
\scalebox{0.26}{\rotatebox{0}{\includegraphics*{4x32_5.264_rev_mom.eps}}}
\caption{Lattice $4\times 32^3$, values of the reversibility estimators.}\label{32_rev}
\end{figure}
\begin{figure}[p]
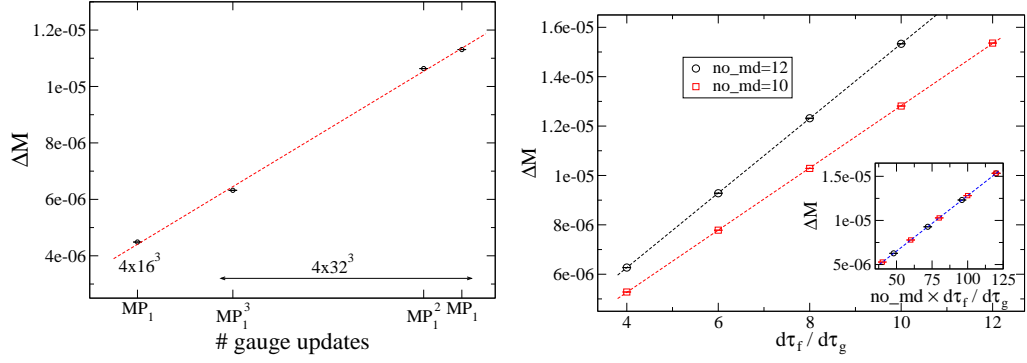

\centering
\scalebox{0.26}{\rotatebox{0}{\includegraphics*{gauge_scaling.eps}}}\hspace{0.2cm}
\scalebox{0.26}{\rotatebox{0}{\includegraphics*{revers.eps}}}
\caption{Scaling of $\Delta M$ with the number of gauge updates. Left: a collection of runs
on different volumes. Right: different runs obtained on a $32^3 \times 4$ lattice and for different
combinations of $n_{md}$ and $d\tau_f/d\tau_g$, reported either as a function
of $d\tau_f/d\tau_g$ or as a function of $n_{md}\times d\tau_f/d\tau_g$. 
 The dashed lines are the result of linear fits.}
\label{gauge_scal_fig}
\end{figure}
\begin{figure}[p]
\centering
\scalebox{0.26}{\rotatebox{0}{\includegraphics*{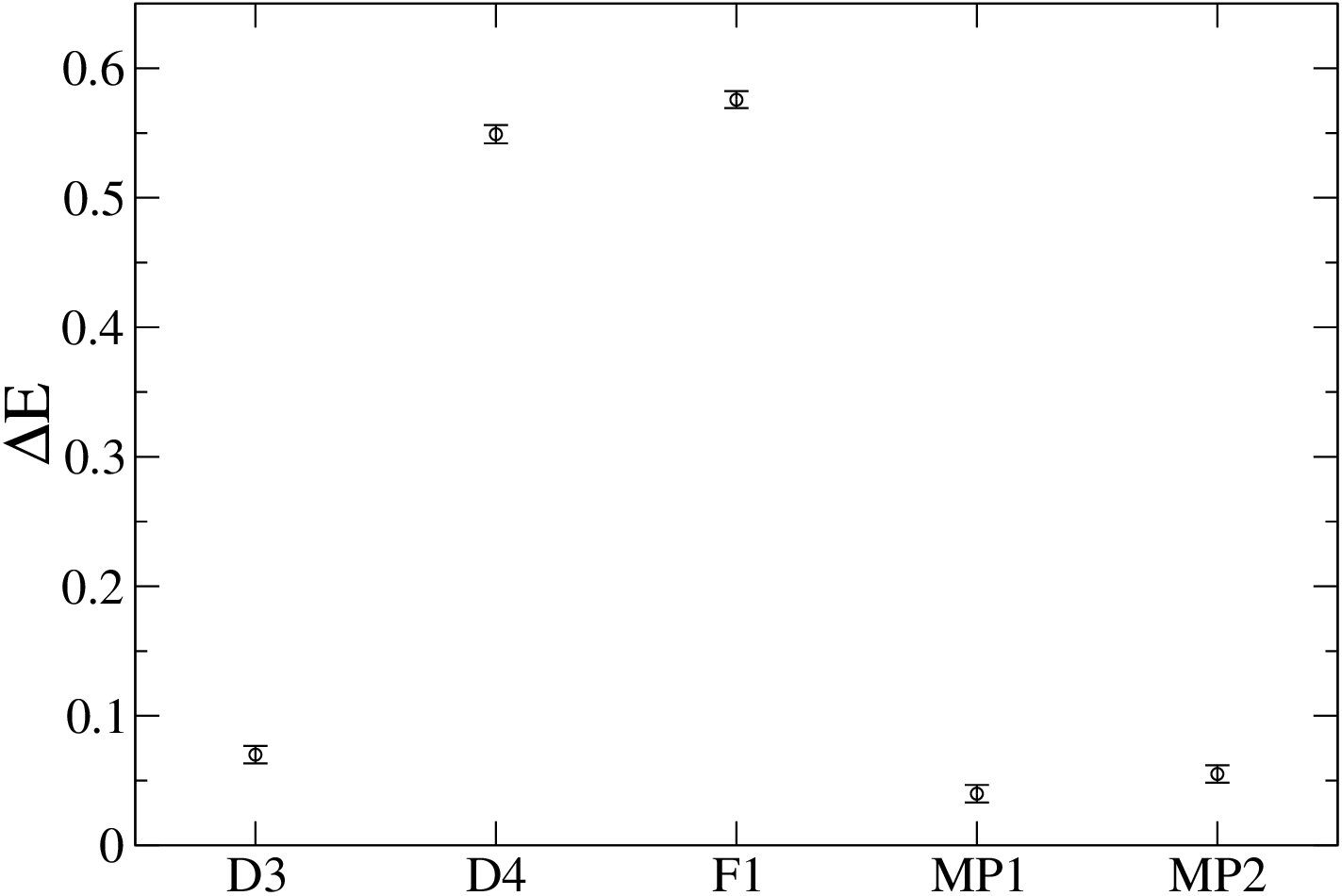}}}
\caption{Lattice $4\times 32^3$, difference between the final and the initial value of the action along an MD trajectory.}
\label{32_dE}
\end{figure}

\subsubsection{$4\times32^3$ lattice}

For this larger lattice we used $n_{md}=16$ and $d\tau_f/d\tau_g=16$. Again all the runs started from a common 
thermalized configuration and the measurement performed in the different runs give compatible results.

Also in this case the reversibility estimators appear to be independent of the precision of the inverter within $1\%$, 
see \Figref{32_rev}, however both estimators are larger than for the $4\times 16^3$ case. Since we have seen that the 
reversibility of the algorithm does not appear to be influenced by the precision of the inverter, it is natural to 
guess that the increased reversibility violation has to be ascribed to the increased number of gauge updates. 
In order to test this guess we have performed other runs, fixing again the total trajectory length to $\tau=1$
and varying $n_{md}$ and $d\tau_f/d\tau_g$; two of them are reported in \Figref{32_rev} as well,
showing reduced violations of reversibility:
\begin{itemize}
\item[$MP_1^{\,2}$] same as $MP_1$ but with $n_{md}=12$ and $d\tau_f/d\tau_g=20$;
\item[$MP_1^{\,3}$] same as $MP_1$ but with $n_{md}=8$ and $d\tau_f/d\tau_g=20$.
\end{itemize}
All results can be summarized by 
\Figref{gauge_scal_fig}, from which it is clear that values obtained for
$\Delta M$ (and analogously for $\Delta C$) for different combinations of 
$n_{md}$ and $d\tau_f/d\tau_g$ collapse on the same linear function when reported
against the number of total gauge updates
(that is $n_{md}\times d\tau_f/d\tau_g$), consistently with the hypothesis that most of the reversibility 
violations is due to the instability generated by the single precision gauge update. 

Although the precision of the inverter does not have large consequences on the reversibility violations, which we just 
saw to be mainly related to the gauge updates, from \Figref{32_dE} it clearly emerges that for large lattices 
double precision is needed in the Metropolis step (\ie in the first and last inversion) in order to have a good 
acceptance ratio. 

The reversibility can be improved by keeping the gauge field and the momenta
in double precision and converting them in single precision only before the force calculation\footnote{We thank one of the referees
for pointing out this simple and effective improvement to us.}. Since in our implementation the momenta are
always stored on the GPU in single precision this is not directly applicable, however just performing the momenta and gauge 
updating\footnote{For the momenta this is just a sum of matrices, for the gauge field it is a matrix product.} in double precision
improves the reversibility violations: for example for the lattice $4\times 32^3$ with the parameter as in the $MP_1$ setting the violations
reduce from
\begin{equation}
\Delta C\sim 4.73\times 10^{-6}\qquad \Delta M\sim 1.13\times 10^{-5}\nonumber
\end{equation}
to
\begin{equation}
\Delta C\sim 8.74\times 10^{-7}\qquad \Delta M\sim 2.08\times 10^{-6}\nonumber
\end{equation}
with a negligible overhead.

\subsection{Performance}

We have compared the performances achieved by our code on C1060 and C2050 architectures
with those obtained by twin codes running on a single CPU core (we have chosen different core architectures) 
and on an apeNEXT crate (256 processors). The twin codes have been reasonably optimized for the respective
architectures, even if room for further optimization may have been left (for instance we have not written
explicit assembly routines for matrix-matrix multiplication).  
The CPU used were an Opteron(tm) 2382 and a Xeon(R) X5560. We have made comparisons on different
lattices $L_s^3 \times 4$ with varying spatial size, and for two different values of the bare quark mass,
$am=0.01335$ and $1.0$; parameters like $n_{md}$ and $d\tau_f/d\tau_g$ have been chosen run by run
to optimize the acceptance ratio. 

\begin{figure}[h]
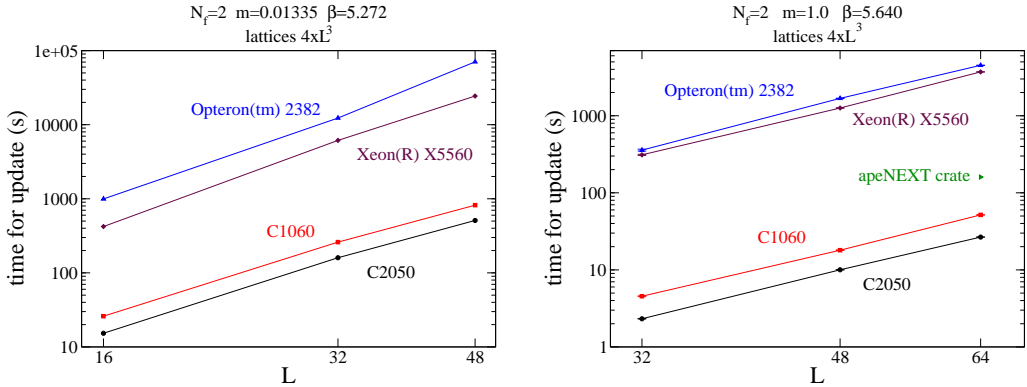

\scalebox{0.26}{\rotatebox{0}{\includegraphics*{times_LM.eps}}}\hspace{0.5cm}
\scalebox{0.26}{\rotatebox{0}{\includegraphics*{times_HM.eps}}}
\caption{Run times on different architectures. For the Opteron and Xeon runs a single core was used.}
\label{time_fig}
\end{figure}

In Fig.~\ref{time_fig} the RHMC update time on different architectures is shown for the different explored
cases. For both the mass values the scaling with the size of the lattice is good. In fact it is a 
characteristic feature of GPUs that increasing the lattice size improves the computational efficiency, as seen also 
in Tab.~\ref{perf_tab}; this happens because with large lattices internal latencies are hidden more effectively. 
Time gains for Tesla C1060 and C2050 are shown in Tab.~\ref{1060_tab} and Tab.~\ref{2050_tab}; particularly impressive 
is the comparison with the results obtained by using an apeNEXT crate. 
Comparing the performance of an $O(100)$ cores GPU with the performance of a single CPU core 
could seem ``unfair'', however in this way we avoid the overhead produced by communications between cores or CPU, which is 
clearly quite large since the algorithm is bandwidth limited. We expect this to balance, at least partially, the effect of more aggressive 
optimizations like the use of inline assembly (which usually improve the performance by an $O(1)$ factor, see e.g. 
\cite{sse}).

\begin{table}[h]
\centering
\begin{tabular}{|l|l|l|l|l|l|l|}
\hline
{ } &  \multicolumn{3}{c}{high mass} \vline & \multicolumn{3}{c}{low mass} \vline \\ 
\hline
spatial size  &  32 & 48 & 64 & 16 & 32 & 48 \\ \hline\hline
Opteron (single core) & 65 & 75 & 75 & 40 & 50 & 85 \\ \hline
Xeon (single core) & 50 & 50 & 50 &  15 & 25 & 30 \\ \hline
apeNEXT crate   & \multicolumn{3}{c}{\(\sim\)3}\vline & \multicolumn{3}{c}{\(\sim\)1}\vline \\ \hline
\end{tabular}
\caption{NVIDIA C1060 time gains over CPU and apeNEXT.} \label{1060_tab}
\end{table} 

\begin{table}[h]
\centering
\begin{tabular}{|l|l|l|l|l|l|l|}
\hline
{ } &  \multicolumn{3}{c}{high mass} \vline & \multicolumn{3}{c}{low mass} \vline \\ 
\hline
spatial size  &  32 & 48 & 64 & 16 & 32 & 48 \\ \hline\hline
Opteron (single core) & 115 & 130 & 140 & 65 & 75 & 140\\ \hline
Xeon (single core) & 85 & 85 & 100 &  30 & 40 & 50 \\ \hline
apeNEXT crate   & \multicolumn{3}{c}{\(\sim\)6}\vline & \multicolumn{3}{c}{\(\sim\)2}\vline \\ \hline
\end{tabular}
\caption{NVIDIA C2050 time gains over CPU and apeNEXT (same code as for C1060, no specific C2050 improvement 
implemented).}\label{2050_tab}
\end{table}

\begin{table}[h]
\centering
\begin{tabular}{|l|l|l|l|l|l|l|}
\hline
{ } &  \multicolumn{3}{c}{C1060} \vline & \multicolumn{3}{c}{C2050} \vline \\ 
\hline
spatial size  &  32 & 48 & 64 & 32 & 48 & 64 \\ \hline\hline
gain factor over Xeon (single core) & 80 & 90 & 120 &  135 & 145 & 210 \\ \hline
\end{tabular}
\caption{NVIDIA C1060 and C2050 time gains over CPU for the pure gauge part sections of the code
(link evolution and pure gauge contribution to momenta evolution in molecular dynamics). 
}\label{hmass_gain}
\end{table}

It is interesting to notice that in the high mass case the time gains over CPU and apeNEXT codes are 
larger with respect to the low mass case, a fact that can be easily explained as follows. 
At low quark mass most of the time is taken by the Dirac matrix inversion, which involves
mostly matrix - vector multiplications: assuming that the on fly reconstruction of the third matrix
row is completely masked, we need to transfer (in single precision) 72 bytes to perform 72 floating point operations, so that
the GPU performance is bandwidth limited to about 100 Gflops. At high quark mass, instead, the effort
needed for Dirac inversion becomes less important and matrix - matrix multiplications needed for the 
pure gauge part of the code take a large fraction of time: in this case we need to transfer 96 bytes (4 rows)
to make 216 floating point operations, so we expect to be roughly a factor 2.25 more efficient than
in the matrix - vector multiplication. In order to test our argument, we have measured separately the 
performances achieved by the pure gauge part of our code, obtaining the time gains reported
in Table ~\ref{hmass_gain}, which, when compared with the low mass performances, 
roughly confirm our estimate; such time gains are
comparable to those obtained by GPU implementations of Monte-Carlo codes for pure gauge theories~\cite{cardoso}.

\section{Further developments: OpenCL and parallelization}
\label{further}

Starting from our single GPU, CUDA implementation of the RHMC code, we are currently developing
new versions of it running on different platforms based on OpenCL and/or on multiGPU architectures.
In this section we report on preliminary results obtained in this direction.

\subsection{Comparison between CUDA and OpenCL implementations}
\label{opencl}

In the latest years increasing interest has grown in the OpenCL project. The main idea of this project is to create a common language for 
any device like CPU, GPU or other accelerators. 

Currently both Nvidia and AMD have a working OpenCL implementation running on 
GPU (Nvidia, AMD) and CPU (AMD); Intel released an OpenCL version for Windows, and other vendors are developing their implementation and 
supporting the project. The idea of developing a single programming language capable of running on different hardware architectures can be a 
great improvement in a sector that evolves very rapidly like the one of GPGPU. 

Starting from our original code running on a single GPU written in Nvidia CUDA, we created a general device independent 
abstraction layer (through the definition of a set of macro) 
which allows to use both CUDA and OpenCL. 
This abstraction layer does not introduce significant overhead: when using Nvidia CUDA the overhead is less than $1\%$, while for 
OpenCL is even less significant, since the OpenCL implementation is lighter than the CUDA one because of the more constrained 
character of the CUDA API.

We have tested our program on the cards C1060, S2050, GT430 (Nvidia) and ATI5870 (ATI) GPUs, by using 
ATI Stream SDK 2.4 and CUDA 3.2. 

\begin{figure}[t]
\centering
\scalebox{0.5}{\rotatebox{0}{\includegraphics*{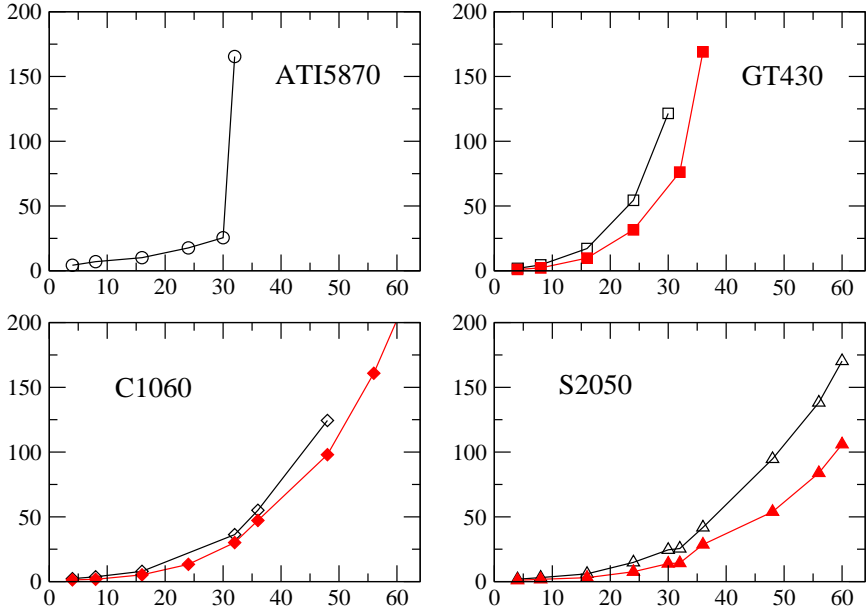}}}
\caption{Update time (in seconds) for a lattice $L_s^3 \times 4$ (the spatial dimension $L_s$ is on the abscissa) for a theory with two fermions of mass 
$m=0.1$ at coupling $\beta=5.59$ and for various GPUs. Results obtained with the OpenCL implementation are shown by empty symbols, while full symbols 
represent the CUDA results.}
\label{cudaopencl_fig}
\end{figure}

A comparison of the efficiency of the two implementations (CUDA and OpenCL) is show in \Figref{cudaopencl_fig}:
we can see that in all the cases where both the CUDA and the OpenCL implementation can be used (\emph{i.e.} on Nvidia GPUs) the CUDA version 
outperforms the OpenCL one. The amount of the performance loss is however strongly hardware dependent: while on the Tesla GPU C1060 we  
observed a $25\%$ performance lose of OpenCL with respect to CUDA, on newer hardware (S2050) this increases to over $60\%$. 

Regarding ATI OpenCL, we have noticed the presence of an increased overhead
for OpenCL API and kernel launching, that reduces the performance on ATI 5870. 
We can see that ATI 5870 can run the
kernel $20\%$ faster than S2050 OpenCL and $50\%$ slower than CUDA Nvidia;
unfortunately we reach the memory bound limit very early and  
we cannot hide big latencies. 

We have also 
tested the OpenCL version of our code when running on multicore CPU architectures, 
obtaining a performance loss of about 2.5 with respect to the single CPU code. That 
means that our present code is not optimized for multi-CPU architectures,
where OpenMP is likely more attractive than OpenCL.

\subsection{Parallelization}
\label{parallel}

Our single GPU version of the code is limited, on available architectures, to 
medium size lattices like $64^3 \times 4$ (finite temperature) or 
$32^4$ (zero temperature). We have shown that on such lattices GPUs
are largely competitive with respect to traditional architectures.
It is then attractive to consider the possibility
of developing a multiGPU version of it, capable of running on a GPU cluster
and thus competitive also for large scale simulations with other 
dedicated parallel machines. At this stage we have only developed
a version capable of running various GPUs connected to the same node,
i.e. we have still not implemented inter-node communications and we 
are thus limited to a small number of GPUs; anyway we can already
make preliminary statements about the scalability of our code.

Parallelization has been based on the abstraction layer described
in \ref{opencl}, which indeed was mainly built 
in order to introduce a multi device abstraction layer, built over OpenCL and
CUDA or any other new technology to come in the future. We have introduced parallelization
by adding first neighbors borders on the fields and updating them when needed. The present
structure of the borders lets us split the lattice only along one direction, 
X, Y, Z or T, because of the need of next-to-nearest neighbors information in the 
computation of the gauge link staple (the standard Dirac kernel can instead be already splitted 
along more than one direction). 

\begin{table}[h]
\centering
\begin{tabular}{|c|c|c|c|}
\hline
lattice size & 1 GPU & 2 GPUs & 4GPUs \\ \hline
$4\times 64^3$ & $239$ &  $134$  &  $95$ \\ \hline
$4\times 96^3$ & $\,\,820^*$ &  $\,\,430^*$  &  $249$ \\ \hline
\end{tabular}
\caption{NVIDIA C1060 update time (in seconds) by using $1$, $2$ or $4$ GPUs (CUDA implementation). The numbers denoted by ${}^*$ are extrapolated 
from simulations performed on smaller lattice sizes because of the impossibility to allocate the corresponding large lattices in the device memory.}
\label{parallel_tab}
\end{table}

In general we have 3 different stages to synchronize the border: we build the buffer border from
the field, we transfer it to the device, then we flush it into the field. In the particular case of parallelization along the 
T direction we can reduce these steps to only the transfer one.
The build and flush stages add an overhead of about $12\%$ on
big lattices (comparable to a $48^3 \times 4 $ on the single GPU)
that reduces slowly when further increasing the lattice size. 
In order to hide the time needed for transfer, we try to overlap it with computation,
and in particular with the shift update inside
the inverter code.

Regarding performances (see Table~\ref{parallel_tab}), on a $64^3 \times 4$ we have an efficiency,
compared to the single GPU case, of $86\%$ on two S2050 (boost $\times 1.73$), of $89\%$ on two C1060 ($\times 1.78$) 
and of $63\%$ ($\times2.51$) on
four C1060; in the last case, increasing the lattice size, we can reach $82\%$ ($\times3.3$) on a $96^3 \times 4$
lattices.
All tests have been performed by splitting along the Y direction and the inefficiencies can be explained by the use of the two additional kernels
needed to align the border before communication.
On large lattices we obtain therefore a good scaling, comparable to what 
reported in Ref.~\cite{multigpu}, which is promising for the extension to the multinode 
implementation at  which we are currently working.
On smaller lattices, instead, the
transfer time cannot be hidden anymore and the boost decreases rapidly.

We have a current new line of development to overcome
the splitting problem, based on building 
the border according to the general topology of a given lattice operator, 
which will permit to compute the border part of an operator separately from the interior part, in order
to overlap not only with the shift update but also with internal computations of
the operator. Such improvement may be particularly useful in the case of improved multi-link
actions and operators and may also introduce a better memory access
pattern.

\section{Conclusions}

The extremely high computation capabilities of modern GPUs make them attractive platforms for high-performance 
computations. Previous studies on lattice QCD applications have been devoted almost exclusively to the Dirac matrix 
inversion problem. We have shown that it is possible to use GPUs to efficiently perform a complete simulation, 
without the need to rely on more traditional architectures: in this case the GPU is not just 
an accelerator, but the real computer.

Our strategy therefore has been that of bringing as much as possible
of the computations on the GPU, leaving for the CPU only light or 
control tasks: in particular the whole molecular dynamics evolution
of gauge fields and momenta, which is the most costly part of 
the Hybrid Monte Carlo algorithm, runs completely on the GPU, thus 
reducing the costly CPU-GPU communications at the minimum.

Following such strategy, we have developed a single GPU code
based on CUDA and tested it on C1060 and C2050 architectures.
We have been able to reach boost factors up to $\sim 10^2$ as compared to 
what reached by a twin traditional C++ code running on a single CPU core. 
Our code is currently in use to study the properties
of strong interactions at finite temperature and density and the 
nature of the deconfinement transition, in particular first 
production results have been reported in Ref.~\cite{rwe2}.

A point worth noting is that in our implementation we have to
rely on the reliability of the GPU. If the GPU is used just for the Dirac matrix inversion the 
result can then be directly checked on the CPU without introducing significant overhead in the computation. 
Such a simple test can not be performed if the GPU is used to perform a complete MD trajectory.
For this reason it was mandatory to use GPUs of the Tesla series.

During the editorial processing of this paper it was signaled us that also the 
TWQCD Collaboration uses GPUs to completely perform the Hybrid Monte Carlo update of QCD with optimal Domain Wall fermions
(their first results were published in \cite{TWQCD}).

Reported performances make surely GPUs the preferred choice
for medium size lattice groups who need enough computational
power at a convenient cost, in this sense they already represent
a breakthrough for the lattice community.
Our current lines of development regard the extension of our code
to OpenCL and to multiGPU architectures and we have reported
preliminary results about that in Section~\ref{further}:
that will open to possibility
to use GPU clusters with fast connection links (see for instance Ref.~\cite{apenet})
in order to make the GPU technology available also for large
scale simulations.

\section*{Acknowledgments}

It's a pleasure to thank A.~Di~Giacomo: without his encouragement and support much of the work reported here 
could not have been realized. 
Test simulations have been run mostly on two GPU farms located in Pisa and Genoa and provided by INFN.
We thank Massimo Bernaschi, Edmondo Orlotti and the APE group in Rome for the possibility
of an early usage of Fermi cards during the first stages of our test runs.
We thank T.-W.~Chiu and M.~A.~Clark for useful comments and one of the referees for his suggestion
to improve the reversibility.

This work is supported in part by the HPCI Strategic Program of
  Ministry of Education. 



\end{document}